\documentclass{tlp}
\submitted{11 April 2006}
\revised{4 October 2006, 17 May 2007}
\accepted{23 May 2007}

\input epsf
\usepackage{oldlfont}

\newcommand{\ignore}[1]{}
\newtheorem{definition}{Definition} 

\begin{document}

\long\def\comment#1{}

\title{Linear Tabling Strategies and Optimizations}

\author[N.F. Zhou, T. Sato, and Y.D. Shen]
{Neng-Fa Zhou \\
CUNY Brooklyn College \& Graduate Center \\
zhou@sci.brooklyn.cuny.edu  
\and Taisuke Sato \\
Tokyo Institute of Technology \\
sato@cs.titech.ac.jp 
\and Yi-Dong Shen \\
State Key Laboratory of Computer Science, Institute of Software, \\
Chinese Academy of Sciences \\
ydshen@ios.ac.cn
}

\pagerange{\pageref{firstpage}--\pageref{lastpage}}
\volume{\textbf{10} (3):}
\jdate{March 2002}
\setcounter{page}{1}
\pubyear{2002}

\maketitle

\label{firstpage}

\begin{abstract}
Recently there has been a growing interest of research in tabling in the logic programming community because of its usefulness in a variety of application domains including program analysis, parsing, deductive databases, theorem proving, model checking, and logic-based probabilistic learning. The main idea of tabling is to memorize the answers to some subgoals and use the answers to resolve subsequent variant subgoals. Early resolution mechanisms proposed for tabling such as OLDT and SLG rely on suspension and resumption of subgoals to compute fixpoints. Recently, the iterative approach named linear tabling has received considerable attention because of its simplicity, ease of implementation, and good space efficiency. Linear tabling is a framework from which different methods can be derived based on the strategies used in handling looping subgoals. One decision concerns when answers are consumed and returned. This paper describes two strategies, namely, {\it lazy} and {\it eager} strategies, and compares them both qualitatively and quantitatively. The results indicate that, while the lazy strategy has good locality and is well suited for finding all solutions, the eager strategy is comparable in speed with the lazy strategy and is well suited for programs with cuts. Linear tabling relies on depth-first iterative deepening rather than suspension to compute fixpoints. Each cluster of inter-dependent subgoals as represented by a top-most looping subgoal is iteratively evaluated until no subgoal in it can produce any new answers. Naive re-evaluation of all looping subgoals, albeit simple, may be computationally unacceptable. In this paper, we also introduce semi-naive optimization, an effective technique employed in bottom-up evaluation of logic programs to avoid redundant joins of answers, into linear tabling. We give the conditions for the technique to be safe (i.e. sound and complete) and propose an optimization technique called {\it early answer promotion} to enhance its effectiveness. Benchmarking in B-Prolog demonstrates that with this optimization linear tabling compares favorably well in speed with the state-of-the-art implementation of SLG.
\end{abstract}
\begin{keywords}
Prolog, Semi-naive evaluation, Recursion, Tabling, Memoization, Linear tabling.
\end{keywords}

\section{Introduction}
The SLD resolution used in Prolog may not be complete or efficient for programs in the presence of recursion. For example, for a recursive definition of the transitive closure of a relation, a query may never terminate under SLD resolution if the program contains left-recursion or the graph represented by the relation contains cycles even if no rule is left-recursive. For a natural definition of the Fibonacci function, the evaluation of a subgoal under SLD resolution spawns an exponential number of subgoals, many of which are variants. The lack of completeness and efficiency in evaluating recursive programs is problematic: novice programmers may lose confidence in writing declarative programs that terminate and real programmers have to reformulate a natural and declarative formulation to avoid these problems, resulting in cluttered programs.

Tabling \cite{Tamaki86,Warren92} is a technique that can get rid of infinite loops for bounded-term-size programs and redundant computations in the execution of recursive programs. The main idea of tabling is to memorize the answers to subgoals and use the answers to resolve their variant descendents. Tabling helps narrow the gap between declarative and procedural readings of logic programs. It not only is useful in the problem domains that motivated its birth, such as program analysis \cite{Dawson96}, parsing \cite{Eisner04,Johnson95,Warren99}, deductive databases \cite{Liu99,Ram95,Sagonas94}, and theorem proving \cite{Nielson04,Pientka03}, but also has been found essential in several other problem domains such as model checking \cite{Ram02} and logic-based probabilistic learning\cite{sato01,Zhou03:prism}. This idea of caching previously calculated solutions, called {\it memoization}, was first used to speed up the evaluation of functions \cite{Michie68}. OLDT \cite{Tamaki86} is the first resolution mechanism that accommodates the idea of tabling in logic programming and XSB is the first Prolog system that successfully supports tabling \cite{Sagonas98}. Tabling has become a practical technique thanks to the availability of large amounts of memory in computers. It has become an embedded feature in a number of other logic programming systems such as B-Prolog \cite{Zhou00,Zhou04}, Mercury \cite{Somogyi06}, TALS \cite{Guo01}, and YAP \cite{Rocha04}. 

OLDT, and SLG \cite{Chen96} alike, is non-linear in the sense that the state of a consumer must be preserved before execution backtracks to its producer. This non-linearity requires freezing stack segments \cite{Sagonas98} or copying stack segments into a different area \cite{Demoen99} before backtracking takes place. Linear tabling is an alternative tabling scheme \cite{Shen01,Zhou00,Zhou03,Zhou04}. The main idea of linear tabling is to use iterative computation of looping subgoals rather than suspension and resumption of them as is done in OLDT to compute fixpoints. This basic idea dates back to the ET* algorithm \cite{Dietrich87}. The DRA method proposed in \cite{Guo01} is based on the same idea but employs different strategies for handling looping subgoals and clauses. In linear tabling, a cluster of inter-dependent subgoals as represented by a {\it top-most looping subgoal} is iteratively evaluated until no subgoal in it can produce any new answers.  Linear tabling is relatively easy to implement on top of a stack machine thanks to its linearity, and is more space efficient than OLDT since the states of subgoals need not be preserved.

Linear tabling is a framework from which different methods can be derived based on the strategies used in handling looping subgoals. One decision concerns when answers are consumed and returned. The {\it lazy} strategy postpones the consumption of answers until no answers can be produced. It is in general space efficient because of its locality and is well suited for all-solution search programs. The {\it eager} strategy, in contrast, prefers answer consumption and return over production. It is well suited for programs with cuts. These two strategies have been compared in SLG-WAM as two scheduling strategies called {\it local} and {\it single-stack} \cite{Freire98}. This paper gives a comprehensive analysis of these two strategies and compares their performance experimentally.

Linear tabling relies on iterative evaluation of top-most looping subgoals to compute fixpoints. Naive re-evaluation of all looping subgoals may be computationally expensive. {\it Semi-naive optimization} is an effective technique used in bottom-up evaluation of Datalog programs \cite{Banc86,Ullman88}. It avoids redundant joins by ensuring that the join of the subgoals in the body of each rule must involve at least one new answer produced in the previous round. The impact of semi-naive optimization on top-down evaluation had been unknown before \cite{Zhou04}. In this paper, we also propose to introduce semi-naive optimization into linear tabling. We have made efforts to properly tailor semi-naive optimization to linear tabling. In our semi-naive optimization, answers for each tabled subgoal are divided into three regions as in bottom-up evaluation, but answers are consumed sequentially until exhaustion not incrementally as in bottom-up evaluation so that answers produced in a round are consumed in the same round. We have found that incremental consumption of answers does not fit linear tabling since it may require more iterations to reach fixpoints. Moreover, consuming answers incrementally may cause redundant consumption of answers. We further propose a technique called {\it early promotion} of answers to reduce redundant consumption of answers.  Our benchmarking shows that this technique gives significant speed-ups to some programs.

An efficient tabling system has been implemented in B-Prolog,\footnote{www.bprolog.com} in which the lazy strategy is employed by default but the eager strategy can be used through declarations for subgoals that are in the scopes of cuts or are not required to return all the answers. Our tabling system not only consumes considerably less stack space than XSB for some programs but also compares favorably well in speed with XSB.

The theoretical framework of linear tabling is given in \cite{Shen01}. The main objective of this paper is to propose evaluation strategies and their optimizations for linear tabling. The remainder of the paper is structured as follows: In the next section we define the terms used in this paper. In Section 3 we give the linear tabling framework and the two answer consumption strategies. In Section 4 we introduce semi-naive optimization into linear tabling and prove its completeness. In Section 5 we describe the implementation of our tabling system and also show how to implement semi-naive optimization. In Section 6 we compare the tabling strategies experimentally, evaluate the effectiveness of semi-naive optimization, and also compare the performance of B-Prolog with XSB. In Section 7 we survey the related work and in Section 8 we conclude the paper.

\section{Preliminaries}
In this section we give the definitions of the terms to make this paper as much self-contained as possible. The reader is referred to \cite{Lloyd88} for a description of SLD resolution. In this paper, we always assume the top-down strategy for selecting clauses and the left-to-right computation rule.

Let $P$ be a program. Tabled predicates in $P$ are explicitly declared and all the other predicates are assumed to be non-tabled. A subgoal of a tabled predicate is called a {\it tabled subgoal}. Tabled predicates are transformed into a form that facilitates execution: each rule ends with a dummy subgoal named $memo(H)$ where $H$ is the head, and each tabled predicate contains a dummy ending rule whose body contains only one subgoal named {\it check\_completion(H)}. For example, given the definition of the transitive closure of a relation,
\begin{verbatim}
   :-table p/2.
   p(X,Y):-p(X,Z),e(Z,Y).
   p(X,Y):-e(X,Y).
\end{verbatim}
The transformed predicate is as follows:
\begin{verbatim}
   p(X,Y):-p(X,Z),e(Z,Y),memo(p(X,Y)). 
   p(X,Y):-e(X,Y),memo(p(X,Y)).               
   p(X,Y):-check_completion(p(X,Y)).          
\end{verbatim}

A table is used to record subgoals and their answers. For each subgoal and its variants, there is an entry in the table that stores the state of the subgoal (e.g., complete or not) and an answer table for holding the answers generated for the subgoal. Initially, the answer table is empty. 

\begin{definition} 
Let $t_1$ and $t_2$ be two terms with no shared variables. The term $t_1$ {\it subsumes} $t_2$ if there exists a substitution $\theta$ such that $t_1\theta$=$t_2$. The two terms $t_1$ and $t_2$ are called {\it variants} if they subsume each other. 
\end{definition}

\begin{definition} 
Let $G=(A_1,A_2,...,A_k)$ be a goal. The first subgoal $A_1$ is called the {\it selected subgoal} of the goal. $G'$ is {\it derived} from $G$ by using a tabled {\it answer} $F$ if there exists a unifier $\theta$ such that $A_1\theta=F$ and $G'=(A_2,...,A_k)\theta$. $G'$ is {\it derived} from $G$ by using a rule ``$H:-B_1,...,B_m$'' if $A_1\theta=H\theta$ and $G'=(B_1,...,B_m,A_2,...,A_k)\theta$. $A_1$ is said to be the {\it parent} of $B_1$, ..., and $B_m$. The relation {\it ancestor} is defined recursively from the parent relation.
\end{definition}

\begin{definition} 
A tabled subgoal that occurs first in the construction of an SLD tree is called a {\it pioneer}, and all subsequent variants are called {\it followers} of the pioneer. Let $G_0$ be a given goal, and $G_0\Rightarrow G_1\Rightarrow\ldots\Rightarrow G_n$ be a {\it derivation} where each goal is derived from the goal immediately preceding it. Let $G_i\Rightarrow \ldots\Rightarrow G_j$ be a sub-sequence of the derivation where $G_i=(A...)$ and $G_j=(A'...)$. The sub-sequence forms a {\it loop} if $A$ and $A'$ are variants. The subgoals $A$ and $A'$ are called {\it looping subgoals}. In particular, $A$ is called the {\it pioneer looping subgoal} and $A'$ is called the {\it follower looping subgoal} of the loop.
\end{definition}

Notice that the pioneer and follower looping subgoals are not required to have the ancestor-descendent relationship, and thus a derivation that contains two variant subgoals may not be a {\it real} loop. Consider, for example, the goal ``$p(X),p(Y)$'' where $p$ is defined by facts. The derivation ``$p(X),p(Y)$'' $\Rightarrow$ $p(Y)$ is treated as a loop although the selected subgoal $p(Y)$ in the second goal is not a descendant of $p(X)$.

\begin{definition}  
A subgoal $A$ is said to be {\it dependent} on another subgoal $A'$ if $A'$ occurs in a derived goal from $A$, i.e., $A\Rightarrow\ldots\Rightarrow(A'...)$. Two subgoals are said to be {\it inter-dependent} if they are dependent on each other. Inter-dependent subgoals constitute a {\it cluster}, which is called a {\it strongly connected component} elsewhere \cite{Sagonas98}. A subgoal in a cluster is called the {\it top-most} subgoal of the cluster if none of its ancestors is included in the cluster.
\end{definition}

\begin{center}
\begin{figure}
\epsfxsize=4cm 
\centering{\epsfbox{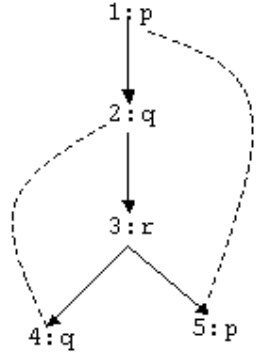}}
\caption{\label{fig:loops}A top-most looping subgoal.}
\end{figure}
\end{center}

Unless a cluster contains only a single subgoal, its top-most subgoal must also be a looping subgoal. For example, the subgoals at the nodes in the SLD tree in Figure \ref{fig:loops} constitute a cluster and the subgoal {\tt p} at node 1 is the top-most looping subgoal of the cluster.

\section{Linear Tabling and Answer Consumption Strategies}
Linear tabling takes a transformed program and a goal, and tries to find a path in the SLD tree that leads to an empty goal. The primitive $table\_start(A)$ is executed when a tabled subgoal $A$ is encountered. Just as in SLD resolution, linear tabling explores the SLD tree in a depth-first fashion, taking special actions when {\it table\_start(A)}, {\it memo(A)}, and {\it check\_completion(A)} are encountered.  Backtracking is done in exactly the same way as in SLD resolution. When the current path reaches a dead end, meaning that no action can be taken on the selected subgoal, execution backtracks to the latest previous goal in the path and continues with an alternative branch. When execution backtracks to the top-most looping subgoal of a cluster, however, we cannot fail the subgoal even after all the alternative clauses have been tried. In general, the evaluation of a top-most looping subgoal must be iterated until its fixpoint is reached. We call each iteration of a top-most looping subgoal a {\it round}.

Various linear tabling methods can be devised based on the framework. A linear tabling method comprises strategies used in the three primitives: {\it table\_start(A)}, {\it memo(A)}, and {\it check\_completion(A)}. In linear tabling, a pioneer subgoal has two roles: one is to produce answers into the table and the other is to return answers to its parent through its variables. Different strategies can be used to produce and return answers. The {\it lazy strategy} gives priority to answer production and the {\it eager strategy} prefers answer consumption over production. In the following we define the three primitives in both strategies.

\subsection{\label{subsection:lazy}The lazy strategy}
The lazy strategy postpones the consumption of answers until no answers can be produced. In concrete, for top-most looping subgoals no answer is returned until they are complete, and for other pioneer subgoals answers are consumed only after all the rules have been tried.

\subsubsection{$table\_start(A)$}
This primitive is executed when a tabled subgoal $A$ is encountered. The subgoal $A$ is registered into the table if it is not registered yet. If $A$'s state is {\it complete} meaning that $A$ has been completely evaluated before, then $A$ is resolved by using the answers in the table. 

If $A$ is a pioneer, meaning that it is encountered for the first time in the current path, then different actions are taken depending on $A$'s state. If $A$'s state is {\it evaluated} meaning that $A$ has occurred before in a different path during the current round, then it is resolved by using answers. Otherwise, if $A$ has never occurred before during the current round, it is resolved by using rules. In this way, a pioneer subgoal needs to be evaluated only once in each round.

If $A$ is a follower of some ancestor $A_0$, meaning that a loop has been encountered,\footnote{As to be discussed later, $A_0$ must be an ancestor of $A$ under the lazy strategy.} then it is resolved by using the answers in the table. After the answers are exhausted, $A$ fails. Failing $A$ is unsafe in general since it may have not returned all of its possible answers. For this reason, the top-most looping subgoal of the cluster of $A$ needs be iterated until no new answer can be produced.

\subsubsection{$memo(A)$}
This primitive is executed when an answer is found for the tabled subgoal $A$. If the answer $A$ is already in the table, then just fail; otherwise fail after the answer is added into the table. The failure of {\it memo} postpones the return of answers until all rules have been tried.

\subsubsection{\label{sub:evaluated}$check\_completion(A)$}
This primitive is executed when the subgoal $A$ is being resolved by using rules and the dummy ending rule is being tried. If $A$ has never occurred in a loop, then $A$'s state is set to {\it complete} and $A$ is failed after all the answers are consumed. 

If $A$ is a top-most looping subgoal, we check if any new answers are produced during the last iteration of the cluster under $A$. If so, $A$ is re-evaluated by calling $table\_start(A)$ after all the dependent subgoals's states are initialized. Otherwise, if no new answer is produced, $A$ is resolved by using answers after its state and all its dependent subgoals' states are set to {\it complete}. Notice that a top-most looping subgoal does not return any answers until it is complete.

If $A$ is a looping subgoal but not a top-most one, $A$ will be resolved by using answers after its state is set to {\it evaluated}. Notice that $A$'s state cannot be set to {\it complete} since $A$ is contained in a loop whose top-most subgoal has not been completely evaluated. For example, in Figure \ref{fig:loops}, {\tt q} reaches its fixpoint only after the top-most looping subgoal {\tt p} reaches its fixpoint. 

As described in the definition of $table\_start(A)$, an {\it evaluated} subgoal is never evaluated using rules again in the same round. This optimization is called {\it subgoal optimization} in \cite{Zhou03}. If evaluating a  subgoal produces some new answers then the top-most looping subgoal will be re-evaluated and so will the subgoal; and if evaluating a subgoal does not produce any new answer, then evaluating it again in the same round would not produce any new answers either. Therefore, the subgoal optimization is safe.

\subsubsection{Example}
Consider the following program, where {\tt p/2} is tabled, and the query {\tt p(a,Y0)}.
\begin{verbatim}
   p(X,Y):-p(X,Z),e(Z,Y),memo(p(X,Y)). (p1)
   p(X,Y):-e(X,Y),memo(p(X,Y)).        (p2)       
   p(X,Y):-check_completion(p(X,Y)).   (p3)

   e(a,b).                             
   e(b,c).                             
\end{verbatim}
The following shows the steps that lead to the production of the first answer:
\begin{tabbing}
aa \= aaa \= aaa \= aaa \= aaa \= aaa \= aaa \kill
\> \> 1: {\tt \underline{p(a,Y0)}} \\
\> \> \> $\Downarrow${\scriptsize apply p1} \\
\> \> 2: {\tt \underline{p(a,Z1)},e(Z1,Y0),memo(p(a,Y0))} \\
\> \> \> {\scriptsize loop found, backtrack to goal 1} \\
\> \> 1: {\tt \underline{p(a,Y0)}} \\
\> \> \> $\Downarrow$ {\scriptsize apply p2} \\
\> \> 3: {\tt \underline{e(a,Y0)},memo(p(a,Y0))} \\
\> \> \> $\Downarrow$ {\scriptsize apply e(a,b)} \\
\> \> 4: {\tt \underline{memo(p(a,b))}} \\
\> \> \> $\Downarrow$ {\scriptsize add answer p(a,b)} 
\end{tabbing}      
After the answer {\tt p(a,b)} is added into the table, {\tt memo(p(a,b))} fails. The failure forces execution to backtrack to {\tt p(a,Y0)}.
\begin{tabbing}
aa \= aaa \= aaa \= aaa \= aaa \= aaa \= aaa \kill
\> \> 1: {\tt \underline{p(a,Y0)}} \\
\> \> \> $\Downarrow$ {\scriptsize apply p3} \\
\> \> 5: {\tt \underline{check\_completion(p(a,Y0))}} 
\end{tabbing}      
Since {\tt p(a,Y0)} is a top-most looping subgoal which has not been completely evaluated yet, {\tt check\_completion(p(a,Y0))} does not consume the answer in the table but instead starts re-evaluation of the subgoal.
\begin{tabbing}
aa \= aaa \= aaa \= aaa \= aaa \= aaa \= aaa \kill
\> \> 1: {\tt \underline{p(a,Y0)}} \\
\> \> \> $\Downarrow${\scriptsize apply p1} \\
\> \> 6: {\tt \underline{p(a,Z1)},e(Z1,Y0),memo(p(a,Y0))} \\
\> \> \> $\Downarrow${\scriptsize use answer p(a,b)} \\
\> \> 7: {\tt \underline{e(b,Y0)},memo(p(a,Y0))} \\
\> \> \> $\Downarrow${\scriptsize apply e(b,c)} \\
\> \> 8: {\tt \underline{memo(p(a,c))}}
\end{tabbing}      
When the follower {\tt p(a,Z1)} is encountered this time, it consumes the answer {\tt p(a,b)}. The current path leads to the second answer {\tt p(a,c)}. On backtracking, the goal numbered 6 becomes the current goal. 
\begin{tabbing}
aa \= aaa \= aaa \= aaa \= aaa \= aaa \= aaa \kill
\> \> 6: {\tt \underline{p(a,Z1)},e(Z1,Y0),memo(p(a,Y0))} \\
\> \> \> $\Downarrow${\scriptsize use answer p(a,c)} \\
\> \> 9: {\tt \underline{e(c,Y0)},memo(p(a,Y0))}
\end{tabbing}
Goal 9 fails. Execution backtracks to the top goal and tries the clause {\tt p3} on it.
\begin{tabbing}
aa \= aaa \= aaa \= aaa \= aaa \= aaa \= aaa \kill
\> \> 1: {\tt \underline{p(a,Y0)}} \\
\> \> \> $\Downarrow$ {\scriptsize apply p3} \\
\> \> 10: {\tt \underline{check\_completion(p(a,Y0))}} 
\end{tabbing}      
Since the new answer {\tt p(a,c)} is produced in the last round, the top-most looping subgoal {\tt p(a,Y0)} needs to be evaluated again. The next round produces no new answer and thus the subgoal's state is set to {\it complete}. After that the top-most subgoal returns the answers {\tt p(a,b)} and {\tt p(a,c)}.

\subsubsection{\label{sec:eager-property}Properties of the lazy strategy}
Under the lazy strategy, answers are not returned immediately after they are produced but are returned via the table after all clauses are tried. No answer is returned for a top-most looping subgoal until the subgoal is complete.

All loops are guaranteed to be real: for any loop $G_i=(A\ldots)\Rightarrow \ldots\Rightarrow G_j=(A'\ldots)$ where $A$ and $A'$ are variants, $A$ must be an ancestor of $A'$. Because each cluster of inter-dependent subgoals is completely evaluated before any answers are returned to outside of the cluster, the lazy strategy has good locality and is thus suited for finding all solutions. For example, when the subgoal $p(Y)$ is encountered in the goal ``{\tt p(X),p(Y)}'', the subtree for {\tt p(X)} must have been explored completely and thus needs not be saved for evaluating {\tt p(Y)}. 

The cut operator cannot be handled efficiently under the lazy strategy. The goal ``$p(X),!,q(X)$'' produces all the answers for $p(X)$ even though only one is needed. 

\subsection{\label{subsection:eager}The eager strategy}
The eager strategy prefers answer consumption and return over production. For a pioneer, answers are used first and rules are used only after all available answers are exhausted, and moreover a new answer is returned to its parent immediately after it is added into the table. The following describes how the three primitives behave under the eager strategy.

\subsubsection{$table\_start(A)$}
Just as in the lazy strategy, $A$ is registered if it is not registered yet. $A$ is resolved by using the tabled answers if $A$ is complete or $A$ is a follower of some former variant subgoal. If $A$ is a pioneer, being encountered for the first time in the current round, it is resolved by using answers first, and then rules after all existing answers are exhausted.

\subsubsection{$memo(A)$}
If the answer $A$ is already in the table, then this primitive fails; otherwise, this primitive succeeds after adding the answer $A$ into the table. Notice that $A$ is returned immediately after it is added into the table. If $A$ is not new, then it must have been returned before.

\subsubsection{$check\_completion(A)$}
If $A$ is a top-most looping subgoal, just as in the lazy strategy, we check whether any new answers are produced during the last iteration of $A$. If so, $A$ is evaluated again by calling $table\_start(A)$. Otherwise, if no new answer is produced, this primitive fails after $A$'s and all its dependent subgoals' states are set to {\it complete}. If $A$ is a looping subgoal but not a top-most one, this primitive fails after $A$'s state is set to {\it evaluated}. An {\it evaluated} subgoal is never evaluated using rules again in the same round. Notice that unlike under the lazy strategy, the primitive $check\_completion(A)$ never returns any answers under the eager strategy. As described above, all the available answers must have been returned by $table\_start(A)$ and $memo(A)$ by the time $check\_completion(A)$ is executed.

\subsubsection{Example}
Because of the need to re-evaluate a top-most looping subgoal, redundant solutions may be observed for a query. Consider, for example, the following program and the query ``{\tt p(X),p(Y)}''.
\begin{verbatim}
   p(1):-memo(p(1)).             (r1)
   p(2):-memo(p(2)).             (r2)       
   p(X):-check_completion(p(X)). (r3)
\end{verbatim}
The following derivation steps lead to the return of the first solution {\tt (1,1)} for {\tt (X,Y)}.
\begin{tabbing}
aa \= aaa \= aaa \= aaa \= aaa \= aaa \= aaa \kill
\> \> 1: {\tt \underline{p(X)},p(Y)} \\
\> \> \> $\Downarrow$ {\scriptsize use r1} \\
\> \> 2: {\tt \underline{memo(p(1))},p(Y)} \\ 
\> \> \> $\Downarrow$ {\scriptsize add answer p(1)} \\
\> \> 3: {\tt \underline{p(Y)}} \\ 
\> \> \> $\Downarrow$ {\scriptsize loop found, use answer p(1)} \\
\end{tabbing}      
When the subgoal {\tt p(Y)} is encountered, it is treated as a follower and is resolved using the tabled answer {\tt p(1)}. After that the first solution {\tt (1,1)} is returned to the top query. When execution backtracks to {\tt p(Y)}, it fails since it is a follower and no more answer is available in the table. Execution backtracks to {\tt p(X)}, which produces and adds the second answer {\tt p(2)} into the table.
\begin{tabbing}
aa \= aaa \= aaa \= aaa \= aaa \= aaa \= aaa \kill
\> \> 1: {\tt \underline{p(X)},p(Y)} \\
\> \> \> $\Downarrow$ {\scriptsize use r2} \\
\> \> 4: {\tt \underline{memo(p(2))},p(Y)} \\ 
\> \> \> $\Downarrow$ {\scriptsize add answer p(2)} \\
\> \> 5: {\tt \underline{p(Y)}} \\ 
\> \> \> $\Downarrow$ {\scriptsize use answer p(1)} \\
\end{tabbing}      
When {\tt p(Y)} is encountered this time, there are two answers {\tt p(1)} and {\tt p(2)} in the table. So the next two solutions returned are {\tt (2,1)} and {\tt (2,2)}. When execution backtracks to goal 1, the dummy ending rule is applied.
\begin{tabbing}
aa \= aaa \= aaa \= aaa \= aaa \= aaa \= aaa \kill
\> \> 1: {\tt \underline{p(X)},p(Y)} \\
\> \> \> $\Downarrow$ {\scriptsize use r3} \\
\> \> 6: {\tt \underline{check\_completion(p(X))},p(Y)} \\
\end{tabbing}      
Since new answers are added into the table during this round, the subgoal {\tt p(X)} needs to be evaluated again, first using answers and then using rules. The second round produces no answer but returns the four solutions {\tt (1,1)}, {\tt (1,2)}, {\tt (2,1)} and {\tt (2,2)} among which only {\tt (1,2)} has not been observed before.

\subsubsection{Properties of the eager strategy}
Since answers are returned eagerly, a pioneer and a follower may not have an ancestor-descendant relationship. Because of the existence of this kind of {\it fake} loops and the necessity of iterating the evaluation of top-most looping subgoals, redundant solutions may be observed. In the previous example, the solutions {\tt (1,1)}, {\tt (2,1)} and {\tt (2,2)} are each observed twice. Provided that the top-most looping subgoal {\tt p(X)} did not return the answer {\tt p(1)} again in the second round, the solution {\tt (1,2)} would have been lost.

The eager strategy is more suited than the lazy strategy for single-solution search. For certain applications such as planning it is unreasonable to find all answers either because the set is infinite or because only one answer is needed. For these applications the eager strategy is more effective than the lazy one. Cuts are handled more efficiently under the eager strategy.

\section{Semi-naive Optimization}
The basic linear tabling framework described in the previous section does not distinguish between new and old answers. The problem with this naive method is that it redundantly joins answers of subgoals that have been joined in early rounds. Semi-naive optimization \cite{Ullman88} reduces the redundancy by ensuring that at least one new answer is involved in the join of the answers for each rule. In this section, we introduce semi-naive optimization into linear tabling and identify sufficient conditions for it to be complete. We also propose a technique called {\it early answer promotion} to further avoid redundant consumption of answers. This optimization works with both the lazy and eager strategies.

\subsection{Preparation}
To make semi-naive optimization possible, we divide the answer table for each tabled subgoal into three regions: 
\begin{center}
\begin{oldtabular}{|c|c|c|} \oldhline
{\it old} & {\it previous} & {\it current} \\ \oldhline
\end{oldtabular}
\end{center}

\noindent
The names of the regions indicate the rounds during which the answers in the regions are produced: {\it old} means that the answers were produced before the previous round, {\it previous} the answers produced during the previous round, and {\it current} the answers produced in the current round. The answers stored in {\it previous} and {\it current} are said to be {\it new}. Before each round is started, answers are promoted accordingly: {\it previous} answers become {\it old} and {\it current} answers become {\it previous}. 

In our optimization, answers are consumed {\it sequentially}. For a subgoal, either all the available answers or only new answers are consumed. This is unlike in bottom-up evaluation where answers are consumed {\it incrementally}, i.e., answers produced in a round are not consumed until the next round. As will be discussed later, incremental consumption of answers as is done in bottom-up evaluation does avoid certain redundant joins but does not fit linear tabling since it may require more rounds to reach fixpoints.

A predicate $p$ {\it calls} a predicate $q$ if: (1) if $q$ occurs in the body of at least one rule in the definition of $p$ ($p$ calls $q$ {\it directly}); or (2) $q$ does not occur in the body of any rule in the definition of $p$ but there exists a predicate in the body of a rule in the definition of $p$ that calls $q$ ($p$ calls $q$ {\it indirectly}). The calling relationship constitutes a graph called a {\it call graph}.

For a given program, we find a level mapping from the predicate symbols in the program to the set of integers to represent the {\it call graph} of the program. Let $m$ be a level mapping. We extend the notation to assume that $m(p(\ldots))=m(p/n)$ for any subgoal $p(\ldots)$ of arity $n$.

\begin{definition} 
{\rm For a given program, a level mapping $m$ represents the {\it call graph} if: for each rule ``$H$$:$$-$$A_1,...,A_n$'' in the program,  $m(H)>m(A_i)$ iff the predicate of $A_i$ does not call (either directly or indirectly) the predicate of $H$, and $m(H)=m(A_i)$ iff the predicates of $H$ and $A_i$ call each other.}
\end{definition}

The level mapping as defined divides predicates in a program into several strata. The predicate at each stratum depends only on those on the lower strata. The level mapping is an abstract representation of the dependence relationship of the subgoals that may occur in execution. If two subgoals $A$ and $A'$ occur in a loop, then it is guaranteed that $m(A)=m(A')$.

\begin{definition} 
{\rm Let ``$H$$:$$-$$A_1,...,A_k,...,A_n$'' be a rule in a program and $m$ be the level mapping that represents the call graph of the program. $A_k$ is called the {\it last depending subgoal} of the rule if $m(A_k)=m(H)$ and $m(H)>m(A_i)$ for $i>k$.}
\end{definition}

The last depending subgoal $A_k$ is the last subgoal in the body that may depend on the head to become complete. Thus, when the rule is re-executed on a subgoal, all the subgoals to the right of $A_k$ that have occurred before must already be complete. 

\begin{definition} 
{\rm Let ``$H$$:$$-$$A_1,...,A_n$'' be a rule in a program and $m$ be a level mapping that represents the call graph of the program. If there is no depending subgoal in the body, i.e., $m(H)>m(A_i)$ for $i=1,...,n$, then the rule is called a {\it base rule}}.
\end{definition}

\subsection{Semi-naive optimization}
\newtheorem{theorem}{Theorem}
\begin{theorem}
Let ``$H$$:$$-$$A_1,...,A_k,...,A_n$'' be a rule where $A_k$ is the last depending tabled subgoal, and $C$ be a subgoal that is being resolved by using the rule in an iteration of a top-most looping subgoal $T$. For a combination of answers of $A_1$, $\cdots$, and $A_{k-1}$, if $C$ has occurred in an early round and the combination does not contain any new answers, then it is safe to let $A_k$ consume new answers only. 
\end{theorem}

\begin{proof}
Because $A_k$ is the last depending subgoal, the subgoals $A_{k+1}$, $\cdots$, and $A_n$ must have been completely evaluated when $C$ is re-evaluated. Let $A_{k_{old}}$ and $A_{k_{new}}$ be the {\it old} and {\it new} answers of the subgoal $A_k$, respectively. For a combination of answers of $A_1$, $\cdots$, and $A_{k-1}$, if the combination does not contain new answers then the join of the combination and $A_{k_{old}}$ must have been done and all possible answers for $C$ that can result from the join must have been produced during the previous round because the subgoal $C$ has been encountered before. Therefore only new answers in $A_{k_{new}}$ should be used.
\end{proof}

\newtheorem{corollary}{Corollary}
\begin{corollary}
Base rules need not be considered in the re-evaluation of any subgoals.
\end{corollary}

Semi-naive optimization would be unsafe if it were applied to new subgoals that have never been encountered before. The following example illustrates this possibility: 

\begin{verbatim}
   ?- p(X,Y).

   :-table p/2.
   p(X,Y) :- p(X,Z),q(Z,Y).  (C1)
   p(b,c) :- p(X,Y).         (C2)
   p(a,b).                   (C3)

   :-table q/2.
   q(c,d) :- p(X,Y),t(X,Y).  (C4)

   t(a,b).                   (C5)
\end{verbatim}

\noindent
In the first round of {\tt p(X,Y)} the answer {\tt p(a,b)} is added to the table by {\tt C3}, and in the second round the rule {\tt C2} produces the answer {\tt p(b,c)} by using the answer produced in the first round. In the third round, the rule  {\tt C1} generates a new subgoal {\tt q(c,Y)} after {\tt p(X,Z)} consumes {\tt p(b,c)}. If semi-naive optimization were applied to {\tt q(c,Y)}, then the subgoal {\tt p(X,Y)} in {\tt C4} could consume only the new answer {\tt p(b,c)} and the third answer {\tt p(b,d)} would be lost.

\subsection{\label{sec:analysis}Analysis}
Semi-naive optimization can lower the complexity of evaluation for some programs. Consider the following example created by David S. Warren:\footnote{Personal communications.}
\begin{verbatim}
   :-table p/2.
   p(X,Y) :- p(X,Z),c(Z,a,Y).
   p(X,Y) :- p(X,Z),c(Z,b,Y).
   p(X,X).
\end{verbatim}
which detects if a given string represented as facts {\tt c}$(I,S,J)$ ($J=I+1$,$S=${\tt a} or $S=${\tt b})  is a sentence of the regular expression $(a|b)^*$. For a string $(ab)^{n/2}$, the query {\tt p(0,$n$)} needs $n/2$ rounds to reach the fixpoint. With semi-naive optimization, the variants of {\tt p(X,Z)} in the bodies consume only new answers, and therefore the program takes linear time. Without semi-naive optimization, however, the program would take $O(n^2)$ time since the variants of {\tt p(X,Z)} would consume all existing answers.

In our semi-naive optimization, answers produced in the current round are consumed immediately rather than postponed to the next round as in the bottom-up version, and answers are promoted each time a new round is started. This way of consuming and promoting answers may cause certain redundancy. 

Consider the conjunction $(P,Q)$. Assume $Q_o$, $Q_p$, and $Q_c$ are the sets of answers in the three regions (respectively, {\it old}, {\it previous}, and {\it current}) of the subgoal $Q$ when $Q$ is encountered in round $i$. Assume also that $P$ had been complete before round $i$ and $P_a$ is the set of answers. The join $P_a \Join (Q_p \bigcup Q_c)$ is computed for the conjunction in round $i$. Assume $Q_o'$, $Q_p'$, and $Q_c'$ are the sets of answers in the three regions when $Q$ is encountered in round i+1. Since answers are promoted before round $i+1$ is started, we have:
\begin{tabbing}
aa \= aaa \= aaa \= aaa \= aaa \= aaa \= aaa \kill
\> $Q_o' = Q_o \bigcup Q_p$ \\
\> $Q_p' = Q_c \bigcup \alpha$ 
\end{tabbing}
where $\alpha$ denotes the new answers produced for $Q$ after the conjunction $(P,Q)$ in round $i$. When the conjunction $(P,Q)$ is encountered in round $i+1$, the following join is computed.
\begin{tabbing}
aa \= aaa \= aaa \= aaa \= aaa \= aaa \= aaa \kill
\> $P_a \Join (Q_p' \bigcup Q_c') = P_a \Join (Q_c \bigcup \alpha \bigcup Qc')$
\end{tabbing}
Notice that the join $P_a \Join Q_c$ is computed in both round $i$ and $i+1$.

We could allow last depending subgoals to consume answers incrementally as is done in bottom-up evaluation, but doing so may require more rounds to reach fixpoints. Consider the following example, which is the same as the one shown above but has a different ordering of clauses:
\begin{verbatim}
   ?- p(X,Y).

   :-table p/2.
   p(a,b).                   (C1)
   p(b,c) :- p(X,Y).         (C2)
   p(X,Y) :- p(X,Z),q(Z,Y).  (C3)

   :-table q/2.
   q(c,d) :- p(X,Y),t(X,Y).  (C4)

   t(a,b).                   (C5)
\end{verbatim}
In the first round, {\tt C1} produces the answer {\tt p(a,b)}. When {\tt C2} is executed, the subgoal in the body cannot consume {\tt p(a,b)} since it is produced in the current round. Similarly, {\tt C3} produces no answer either. In the second round, {\tt p(a,b)} is moved to the {\it previous} region, and thus can be consumed. {\tt C2} produces a new answer {\tt p(b,c)}. When {\tt C3} is executed, no answer is produced since {\tt p(b,c)} cannot be consumed. In the third round, {\tt p(a,b)} is moved to the {\it old} region, and {\tt p(b,c)} is moved to the {\it previous} region. {\tt C3} produces the third answer {\tt p(b,d)}. The fourth round produces no new answer and confirms the completion of the computation. So in total four rounds are needed to compute the fixpoint. If answers produced in the current round are consumed in the same round, then only two rounds are needed to reach the fixpoint.

\vspace*{0.5cm}
\subsection{Early promotion of answers}
As discussed above, sequential consumption of answers may cause redundant joins. In this subsection, we propose a technique called {\it early promotion} of answers to reduce the redundancy.

\begin{definition} 
{\rm Let $Q$ be the first follower that exhausts its answers in the current round.
Then all the answers of $Q$ in the {\it current} region are promoted to the {\it previous} region once being consumed by $Q$.}
\end{definition}

Consider again the conjunction $(P,Q)$ where $Q$ is the first follower that exhausts its answers. The answers in the current region $Q_c$ are promoted to the {\it previous} region after $Q$ has consumed all its answers in round $i$. By doing so, the join $P_a \Join Q_c$ will not be recomputed in round $i+1$ since $Q_c$ must have been promoted to the {\it old} region in round $i+1$.

Consider, for example, the following program:
\begin{verbatim}
   ?- p(X,Y).

   :-table p/2.
   p(a,b).                   (C1)
   p(b,c) :- p(X,Y).         (C2)
\end{verbatim}
Before {\tt C2} is executed in the first round, {\tt p(a,b)} is in the {\it current} region. Executing {\tt C2} produces the second answer {\tt p(b,c)}. Since the subgoal {\tt p(X,Y)} in {\tt C2} is the first follower that exhausts its answers in the current round, it is qualified to promote its answers. So the answers {\tt p(a,b)} and {\tt p(b,c)} are moved from the {\it current} region to the {\it previous} region immediately after being consumed by {\tt p(X,Y)}. As a result, the potential redundant consumption of these answers by {\tt p(X,Y)} is avoided in the second round since they will all be transferred to the {\it old} region before the second round starts.

\begin{theorem}
Early promotion does not lose any answers. 
\end{theorem}

\begin{proof}
First note that although answers are tabled in three disjoint regions, all tabled answers will be consumed except for some last depending subgoals that would skip the answers in their {\it old} regions (see Theorem 1). Assume, on the contrary, that applying early promotion loses answers. Then there must be a last depending subgoal $A_k$ in a rule ``$H$$:$$-$$A_1,...,A_k,...,A_n$'' and a tabled answer $A$ for $A_k$ such that $A$ has been moved to the {\it old} region before being consumed by $A_k$ so that $A$ will never be consumed by $A_k$. Assume $A$ is produced in round $i$ by a variant of $A_k$. We distinguish between the following two cases:

\begin{enumerate}
\item The last depending subgoal $A_k$ is not selected in round $i$. In round $j (j>i)$, $A_k$ is selected either because $H$ is new or some $A_s (s<k)$ consumes a new answer. By Theorem 1, $A_k$ will consume all answers in the three regions, including the answer $A$.

\item Otherwise, $A$ must be produced by $A_k$ itself or a variant subgoal of $A_k$ that is selected either {\it before} or {\it after} $A_k$ in round $i$. If $A$ is produced by $A_k$ itself or {\it before} $A_k$ is selected, then the answer will be consumed by $A_k$ since promoted answers will remain new by the end of the round. If $A$ is produced by a variant {\it after} $A_k$ is selected, then the answer cannot be promoted because $A_k$ exhausts its answers {\it before} the variant. In this case, the answer $A$ will remain new in the next round and will thus be consumed by $A_k$.
\end{enumerate}

Both of the above two cases contradict our assumption. The proof then concludes.
\end{proof}

\section{Implementation}
Changes to the Prolog machine ATOAM \cite{Zhou96}  are needed to implement linear tabling. In this section we describe the changes to the data structures and the instruction set. 
\ignore{We also show how to implement the cut operator. }
To make the paper self-contained, we first give an overview of the ATOAM architecture. 

\subsection{An overview of ATOAM}
The ATOAM uses all the data areas used by the WAM. The {\it heap} stores terms created during execution. The register {\tt H} points to the top of the heap. The {\it trail} stack stores updates that must be undone upon backtracking. The register {\tt T} points to the top of the trail stack. The {\it control} stack stores frames associated with predicate calls. 

Unlike in the WAM where arguments are passed through argument registers, arguments in the ATOAM are passed through stack frames and only one frame is used for each predicate call. Each time a predicate is invoked by a call, a frame is placed on top of the local stack unless the frame currently at the top can be reused. Frames for different types of predicates have different structures. For standard Prolog, a frame is either {\it determinate} or {\it nondeterminate}. A nondeterminate frame is also called a {\it choice point}. The register {\tt AR} points to the current frame and the register {\tt B} points to the latest {\it choice point}.

A determinate frame has the following structure:
\begin{center}
\begin{oldtabular}{|l|l|} \oldhline
{\tt A1..An} &  Arguments \\  \oldhline
{\tt AR}     &  Pointer to the parent frame \\  \oldhline
{\tt CP}     &  Continuation program pointer \\  \oldhline
{\tt BTM}    &  Bottom of the frame \\  \oldhline
{\tt TOP}    &  Top of the frame \\  \oldhline
{\tt Y1..Ym} & Local variables  \\ \oldhline
\end{oldtabular}
\end{center}
Where {\tt BTM} points to the bottom of the frame, i.e., the slot for the first argument, and {\tt TOP} points to the top of the frame, i.e., the slot just next to that for the last local variable\footnote{It is a convention in the literature that the stack is assumed to grow downwards}. The {\tt TOP} register points to the next available slot on the stack. The {\tt BTM} slot is not in the original version \cite{Zhou96}. This slot is introduced for supporting garbage collection and co-routining. The {\tt AR} register points to the {\tt AR} slot of the current frame. Arguments and local variables are accessed through offsets with respect to the {\tt AR} slot. An argument or a local variable is denoted as {\tt y(I)} where {\tt I} is the offset. Arguments have positive offsets and local variables have negative offsets. It is the caller's job to place the arguments and fill in the {\tt AR}, and {\tt CP} slots. The callee fills in the {\tt BTM} and {\tt TOP} slots and initializes the local variables.

A choice point frame contains, in addition to the slots in a determinate frame, four slots located between the {\tt TOP} slot and local variables: 
\begin{center}
\begin{oldtabular}{|l|l|} \oldhline
{\tt CPF} &  Backtracking program pointer \\ \oldhline
{\tt H}   &  Top of the heap \\ \oldhline
{\tt T}   &  Top of the trail \\ \oldhline
{\tt B}   &  Parent choice point \\ \oldhline
\end{oldtabular}
\end{center}
The {\tt CPF} slot stores the program pointer to continue with when the current branch fails. The slot {\tt H} points to the top of the heap when the frame is allocated. As in the WAM, a new register, called {\tt HB}, is used as an alias for {\tt B->H}. When a variable is bound, it must be trailed if it is older than {\tt B} or {\tt HB}.

\subsection{The extension of ATOAM for tabling}
A new data area, called {\it table area}, is introduced for memorizing tabled subgoals and their answers. The {\it subgoal table} is a hash table that stores all the tabled subgoals encountered in execution. For each tabled subgoal and its variants, there is an entry in the table, which is a record containing the following information:
\begin{center}
\begin{oldtabular}{|l|} \oldhline
{\tt SubgoalCopy}  \\ \oldhline
{\tt PioneerAR} \\ \oldhline
{\tt State} \\ \oldhline
{\tt TopMostLoopingSubgoal} \\ \oldhline
{\tt DependentSubgoals}  \\ \oldhline
{\tt AnswerTable} \\ \oldhline
\end{oldtabular}
\end{center}

\noindent
The field {\tt SubgoalCopy} points to the copy of the subgoal in the table area. In the copy, all variables are numbered. Therefore all variants of the subgoal are identical. 

The field {\tt PioneerAR} points to the frame of the pioneer, which is needed for implementing cuts. When the choice point of a tabled subgoal is cut off before the subgoal reaches completion, the field {\tt PioneerAR} will be set to {\tt NULL}. When a variant of the subgoal is encountered again after, the subgoal will be treated as a pioneer.

The field {\tt State} indicates whether the subgoal is a looping subgoal, whether the answer table has been revised, and whether the subgoal is {\it complete} or {\it evaluated}. When execution backtracks to a top-most looping subgoal, if the {\it revised} bit is set, then another round will be started for the subgoal. A top-most looping subgoal becomes complete if this {\it revised} bit is unset after a round. At that time, the subgoal and all of its dependent subgoals will be set to {\it complete}. As described in \ref{sub:evaluated}, an {\it evaluated} subgoal is never evaluated again using rules in each round.

The {\tt TopMostLoopingSubgoal} field points to the entry for the top-most looping subgoal, and the field {\tt DependentSubgoals} stores the list of subgoals on which this subgoal depends. When a top-most looping subgoal becomes complete, all of its dependent subgoals turn to complete too. 

The field {\tt AnswerTable} points to the answer table for this subgoal, which is also a hash table. Hash tables expand dynamically. Let {\tt g} be the pointer to the record for a subgoal in the table. The first answer in the answer table is referenced as \verb+g->AnswerTable->FirstAnswer+ and the last answer is referenced as \verb+g->AnswerTable->LastAnswer+. In the beginning, the answer table is empty and both {\tt FirstAnswer} and {\tt LastAnswer} reference a dummy answer.

The frame for a tabled predicate contains the following two slots in addition to those slots stored in a choice point frame:   
\begin{center}
\begin{oldtabular}{|l|}  \oldhline
{\tt SubgoalTable} \\ \oldhline
{\tt CurrentAnswer} \\ \oldhline
\end{oldtabular}
\end{center}

\noindent
The {\tt SubgoalTable} points to the subgoal table entry, and the {\tt CurrentAnswer} points to the last answer that has been consumed. The next answer can be reached from this reference on backtracking. When a frame is created, the slot {\tt CurrentAnswer} is initialized to be \verb+g->AnswerTable->FirstAnswer+ where {\tt g} is the pointer to the record for the tabled subgoal.

Three new instructions, namely {\tt table\_start}, {\tt memo}, and {\tt check\_completion}, are introduced into the ATOAM for encoding the three table primitives. Figure \ref{fig:ins} shows the compiled code of an example program.
\begin{figure}
\begin{verbatim}
  % :-tabled p/2.
  % p(X,Y):-p(X,Z),e(Z,Y).
  % p(X,Y):-e(X,Y).
  p/2: table_start 2,1
       fork r2
       para_value y(2)
       para_var y(-13)
       call p/2          % p(X,Z)
       para_value y(-13)
       para_value y(1)
       call e/2          % e(Z,Y)
       memo
  r2:  fork r3
       para_value y(2)
       para_value y(1)
       call e/2          % e(X,Y)
       memo
  r3:  check_completion p/2
\end{verbatim}
\caption{\label{fig:ins}Compiled code of an example program.}
\end{figure}

The {\tt table\_start} instruction takes two operands: the arity ({\tt 2}) and the number of local variables ({\tt 1}). The {\tt fork} instruction sets the {\tt CPF} slot to hold the address to backtrack to on failure. The parameter passing instructions ({\tt para\_value} and {\tt para\_var} in this example) pass arguments to the callee's frame. The {\tt memo} instruction is executed after an answer has been found. The {\tt check\_completion} instruction takes the entrance ({\tt p/2}) as an operand so that the predicate can be re-entered when it needs re-evaluation.

\subsection{Implementing semi-naive optimization}
To implement semi-naive optimization, we add the following two pointers into the record for each tabled subgoal:
\begin{center}
\begin{oldtabular}{|l|}  \oldhline
{\tt LastOldAnswer} \\ \oldhline
{\tt LastPrevAnswer} \\ \oldhline
\end{oldtabular}
\end{center}
where the pointer {\tt LastOldAnswer} points to the last answer in the old region and the pointer {\tt LastPrevAnswer} points to the last answer in the previous region. The {\tt check\_completion} instruction resets the pointers for all the tabled subgoals in the current cluster before it starts the next round:
\begin{verbatim}
   for each subgoal g in the current cluster {
       g->LastOldAnswer = g->LastPrevAnswer;
       g->LastPrevAnswer = g->AnswerTable->LastAnswer;
   }
\end{verbatim}

The {\tt memo} instruction is changed so that early promotion of answers is performed if the condition for promotion is met. Let {\tt g} be the pointer to the tabled subgoal. If the subgoal has exhausted all its answers in the table and early promotion has never be done before on the subgoal in the same round, then answers in the current region are promoted to the previous region:
\begin{verbatim}
   g->LastPrevAnswer = g->AnswerTable->LastAnswer
\end{verbatim}
The promoted answers will be moved to the old region before the start of the next round.

A bit vector is added into the frame for each tabled predicate to indicate if any new answer has been consumed by any tabled subgoal. Semi-naive optimization can be applied only if no tabled subgoal in the predicate has consumed any new answer.

A new instruction, called {\tt last\_depending\_tabled\_call}, is introduced to encode last depending tabled subgoals.  In the example shown in Figure \ref{fig:ins}, the ``{\tt call p/2}'' instruction is changed to ``{\tt last\_depending\_tabled\_call p/2}'' to enable semi-naive optimization. The {\tt last\_depending\_tabled\_call} instruction has the same behavior as the {\tt call} instruction, but the callee can check the type of the instruction to see if it is invoked by a last depending tabled subgoal.

Let {\tt g} be the pointer to the current tabled subgoal. The {\tt table\_start} instruction sets the {\tt CurrentAnswer} slot of the frame to \verb+g->LastOldAnswer+ so that the subgoal consumes only new answers if: (1) the parent frame is a tabled frame; (2) no bit in the bit vector in the parent frame is set, which means that no tabled subgoal has consumed any new answer; and (3) the predicate is invoked by a {\tt last\_depending\_tabled\_call} instruction. If any of these condition is not satisfied, the {\tt CurrentAnswer} slot is set to \verb+g->AnswerTable->FirstAnswer+ and all the answers will be consumed by the subgoal.

\section{Performance Evaluation}
We empirically compared the two answer consumption strategies and evaluated the effectiveness of semi-naive optimization. We also compared the performance of B-Prolog (version 6.9) with XSB (version 3.0). A Linux machine with 750MHz Intel process and 512GB RAM was used in the experiment. Benchmarks from three different sources were used:\footnote{The benchmarks are available from {\it probp.com/bench.tar.gz}.} Datalog programs shown in Figure \ref{fig:datalog} with randomly generated graphs; the CHAT benchmark suite \cite{Demoen99}; and a parser, called {\it atr}, for the Japanese language defined by a grammar of over 860 rules \cite{Uratani94}.  This section presents the experimental results and reports the statistics to support the results. This section also gives experimental results on the Warren's example for which SLG as implemented in XSB has lower time complexity than linear tabling when semi-naive optimization ceases to be effective.

\begin{figure}[t]
\begin{center}
\begin{tabbing}
aa \= aaa \= aaa \= aaa \= aaa \= aaa \= aaa \kill
tcl: \> \> {\tt tcl(X,Y):-edge(X,Y).} \\
\> \> {\tt tcl(X,Y):-tcl(X,Z),edge(Z,Y).} \\
\\
tcr: \> \> {\tt tcr(X,Y):-edge(X,Y).} \\
\> \> {\tt tcr(X,Y):-edge(X,Z),tcr(Z,Y).} \\
\\
tcn: \> \> {\tt tcn(X,Y):-edge(X,Y).} \\
\> \> {\tt tcn(X,Y):-tcn(X,Z),tcn(Z,Y).} \\
\\
sg: \> \> {\tt sg(X,X).} \\
\> \> {\tt sg(X,Y):-edge(X,XX),sg(XX,YY),edge(Y,YY).} 
\end{tabbing}
\end{center}
\caption{\label{fig:datalog}Datalog programs.}
\end{figure}

\subsection{Comparison of the two answer-consumption strategies}
Table \ref{tab:strategies} compares the two answer-consumption strategies in terms of speed and stack space\footnote{The total usage of the local, global and trail stacks.} efficiencies. The difference of these two strategies in terms of CPU time is small on average. This result implies that for programs with cuts declaring the use of the eager strategy would not cause significant slow-down. The difference in the usage of stack space is more significant than in CPU time. This is because, as discussed before, the lazy strategy has better locality than the eager strategy.

\begin{table}
\begin{small}
\begin{center}
\caption{\label{tab:strategies}Comparison of the lazy and eager strategies.}
\begin{oldtabular}{|c|l|l|l|l|} \oldhline
program  & \multicolumn{2}{c|} {CPU time} & \multicolumn{2}{c|} {Stack space} \\ \cline{2-5} 
         & Lazy & Eager & Lazy & Eager \\ \oldhline \oldhline
tcl      & 1  & 1.02 & 1   & 1.00 \\
tcr      & 1 & 0.96 & 1  & 1.00 \\
tcn      & 1  & 0.90 & 1  & 1.00 \\
sg      & 1 & 0.89 & 1  & 1.02 \\ 
cs\_o    & 1  & 1.17 & 1 & 1.36 \\
cs\_r    & 1  & 1.09 & 1  & 1.36 \\
disj     & 1  & 1.06 & 1  & 1.41 \\
gabriel  & 1  & 1.08 & 1  & 1.18 \\
kalah    & 1  & 1.17 & 1  & 2.03 \\
pg       & 1  & 2.28 & 1  & 3.59 \\
peep     & 1  & 0.99 & 1  & 2.88 \\
read     & 1  & 0.85 & 1  & 2.22 \\
atr      & 1  & 1.03 & 1  & 1.06 \\ \oldhline
$average$& 1          & 1.12 & 1  & 1.62 \\ \oldhline \oldhline
\end{oldtabular}
\end{center}
\end{small}
\end{table}

\subsection{Effectiveness of semi-naive optimization}
Table \ref{tab:semi} shows the effectiveness of semi-naive optimization in gaining speed-ups under both strategies. Without this optimization, the system would consume over 30\% more CPU time on average under either strategy. Our experiment also indicates that on average over 95\% of the gains in speed are attributed to the {\it early promotion} technique.

\begin{table}
\begin{small}
\begin{center}
\caption{\label{tab:semi}Effectiveness of semi-naive optimization.}
\begin{oldtabular}{|c|r|r|} \oldhline
program   &   \multicolumn{2}{c|} {CPU time ($\frac{no semi}{semi}$)} \\ \cline{2-3} 
          &   Lazy & Eager \\ \oldhline \oldhline
tcl      &            2.00 & 1.89 \\
tcr      &            1.22 & 1.19 \\
tcn      &            1.68 & 1.74 \\
sg      &            1.22 & 1.51 \\ 
cs\_o    &            1.10 & 1.10 \\
cs\_r    &            1.09 & 1.10 \\
disj     &            1.52 & 1.46 \\
gabriel  &            1.32 & 1.15 \\
kalah    &            1.52 & 1.41 \\
pg       &            1.21 & 1.05 \\
peep     &            1.09 & 1.11 \\
read     &            1.98 & 1.27 \\
atr      &            1.00 & 1.00 \\ \oldhline 
$average$&            1.38 & 1.31 \\ \oldhline \oldhline
\end{oldtabular}
\end{center}
\end{small}
\end{table}

\subsection{Comparison with XSB}
Table \ref{tab:xsb} compares BP with XSB on time and stack space efficiencies. For XSB, the stack space is the total of the maximum amounts of global, local, trail, choice point, and SLG completion stack spaces. The default setting, namely, the SLG-WAM and the local scheduling strategy, is used. BP is faster than XSB on the Datalog programs and the parser but slower than XSB on the CHAT benchmark suite; and BP consumes considerably less stack space than XSB on some of the programs ({\it tcr}, {\it tcn}, {\it sg}, and {\it atr}).

The results must be interpreted with two differences of the two compared systems taken into account: On the one hand, BP is on average more than twice as fast as XSB for standard Prolog programs, and on the other hand the trie data structure used in XSB \cite{Ram98} is far more advanced than hash tables used in BP for managing the table area. It is unclear to what extent each difference contributes to the overall efficiency.

The YAP implementation of SLG-WAM is up to twice as fast as XSB \cite{Somogyi06} on the transitive closure and same-generation benchmarks with both chain and cyclic graphs. This entails that the BP implementation of linear tabling is comparable in speed with the most sophisticated implementation of SLG-WAM for the Datalog benchmarks.

\begin{table}
\begin{small}
\begin{center}
\caption{\label{tab:xsb}Comparison of B-Prolog and XSB.}
\begin{oldtabular}{|c|c|r|r|} \oldhline
program  &  BP  & \multicolumn{2}{c|} {XSB} \\ \cline{3-4} 
         & (Lazy)  & CPU time & Stack space \\ \oldhline \oldhline
tcl      &            1 & 1.85 & 0.81 \\
tcr      &            1 & 1.46 & 33.41 \\
tcn      &            1 & 1.31 & 32.84\\
sg      &            1 & 1.47 & 109.12\\ 
cs\_o    &            1 & 0.37 & 0.57 \\
cs\_r    &            1 & 0.35 & 0.73 \\
disj     &            1 & 0.68 & 0.82 \\
gabriel  &            1 & 0.61 & 2.05 \\
kalah    &            1 & 1.00 & 0.58 \\
pg       &            1 & 0.76 & 1.85 \\
peep     &            1 & 0.37 & 2.97 \\
read     &            1 & 0.69 & 11.12\\ 
atr      &            1 & 2.26 & 21.24 \\ \oldhline 
\end{oldtabular}
\end{center}
\end{small}
\end{table}

The empirical data on the usage of table space are not reported. BP constantly consumes less table space than XSB for the benchmarks. In BP, both subgoal and answer tables are maintained as dynamic hashtables. In XSB, in contrast, tables are maintained as tries \cite{Ram98}. The usage of table space is independent of the strategies and optimizations. Both BP and XSB would consume the same amount of table space if the same data structure were employed.

\subsection{Statistics on iterations}
Table \ref{tab:its} reports the statistics on the maximum (max its.) and average (ave. its.) numbers of iterations for tabled subgoals to reach their fixpoints.\footnote{Each subgoal has a counter which is initialized when the subgoal is tabled and is incremented each time the subgoal is resolved using rules. Note that semi-naive optimization may reduce the work of each iteration but has no effect on the number of iterations needed to reach the fixpoint.} The column \#subgoals shows the number of tabled subgoals. While for some programs, the maximum number of iterations performed is high (e.g., the maximum number for {\it atr} is 6), the average numbers are quite low. 

The necessity of re-evaluating looping subgoals has been blamed for the low speed of iteration-based tabling systems \cite{Zhou00,Guo01}. Our new findings indicate that re-evaluation is not a dominant factor for the benchmarks. This statistics well explain why an implementation of linear tabling could achieve comparable speed performance with SLG-WAM for the benchmarks.

\begin{table}
\begin{small}
\begin{center}
\caption{\label{tab:its}Statistics on iterations.}
\begin{oldtabular}{|c|r|r|r|} \oldhline
program  &  \#subgoals & max its. & ave. its. \\ \oldhline \oldhline
tcl      &            1 & 2 & 2.00  \\
tcr      &            51 & 2 & 1.96 \\
tcn      &            51 & 2 & 1.98 \\
sg      &             153 & 2 & 1.32 \\ 
cs\_o    &            76 & 2 & 1.14 \\
cs\_r    &            76 & 2 & 1.16 \\
disj     &            74 & 2 & 1.20 \\
gabriel  &            59 & 2 & 1.20 \\
kalah    &            102 & 3 & 1.24 \\
pg       &            48 & 2 & 1.13 \\
peep     &            49 & 3 & 1.29 \\ 
read     &            131 & 5 & 1.34 \\
atr      &            7139 & 6 & 1.81 \\ \oldhline \oldhline
\end{oldtabular}
\end{center}
\end{small}
\end{table}

\subsection{The complexity issue}
The following is a slightly changed version of the Warren's example which disenables semi-naive optimization:
\begin{verbatim}
   :-table p/2.
   p(X,Y) :- q(X,Z),c(Z,a,Y).
   p(X,Y) :- q(X,Z),c(Z,b,Y).
   p(X,X).

   q(X,Y) :- p(X,Y).
\end{verbatim}
Since the last depending subgoals {\tt q(X,Z)} in {\tt p/2} are not tabled, semi-naive optimization cannot be applied to {\tt p/2}.
For a string $(ab)^{n/2}$, the query {\tt p(0,$n$)} needs $n/2$ iterations to reach the fixpoint. Since in each iteration the subgoal {\tt q(X,Z)} is rewritten into {\tt p(X,Z)} which returns all existing answers, the total time taken is $O(n^2)$. In contrast, the program takes only $O(n)$ time under SLG. For the size n=5000, it took BP 3.5 seconds to run the program while XSB only 15 milliseconds. For the original version of the program to which semi-naive optimization is applicable, it took BP only 7 milliseconds.

\section{Related Work}
There are three different tabling schemes, namely OLDT and SLG \cite{Tamaki86,Sagonas98}, CAT \cite{Demoen98,Somogyi06}, and iteration-based tabling including linear tabling \cite{Shen99,Shen01,Zhou00,Zhou03,Zhou04} and DRA \cite{Guo01}. SLG \cite{Chen96} is a formalization based on OLDT for computing well-founded semantics for general programs with negation. The basic idea of using iterative deepening to compute fixpoints dates back to the ET* algorithm \cite{Dietrich87}.

In SLG-WAM, a consumer fails after it exhausts all the existing answers and its state is preserved by freezing the stack so that it can be reactivated after new answers are generated. The CAT approach does not freeze the stack but instead copies the stack segments between the consumer and its producer into a separate area so that backtracking can be done normally. The saved state is reinstalled after a new answer is generated. CHAT \cite{Demoen99} is a hybrid approach that combines SLG-WAM and CAT. 

Linear tabling relies on iterative computation of looping subgoals to compute fixpoints.  Linear tabling is probably the easiest scheme to implement since no effort is needed to preserve states of consumers and the garbage collector can be kept untouched for tabling. Linear tabling is also the most space-efficient scheme since no extra space is needed to save states of consumers. Nevertheless, linear tabling without optimization could be computationally more expensive than the other two schemes. 

The DRA method \cite{Guo01} is also iteration based, but it identifies looping clauses dynamically and iterates the execution of looping clauses to compute fixpoints. While in linear tabling iteration is performed on only top-most looping subgoals, in DRA iteration is performed on every looping subgoal. In ET* \cite{Dietrich87}, every tabled subgoal is iterated even if it does not occur in a loop. Besides the difference in answer consumption strategies and optimizations, the linear tabling scheme described in this paper differs from the original version \cite{Zhou00,Shen01} in that followers fail after they exhaust their answers rather than steal their pioneers' choice points. This strategy is originally adopted in the DRA method.

The two consumption strategies have been compared in XSB \cite{Freire98} as two scheduling strategies. The lazy strategy is called {\it local scheduling} and the eager strategy is called {\it single-stack scheduling}. Another strategy, called {\it batched scheduling}, is similar to local scheduling but top-most looping subgoals do not have to wait until their clusters become complete to return answers. Their experimental results indicate that local scheduling constantly outperforms the other two strategies on stack space and can perform asymptotically better than the other two strategies on speed. The superior performance of local scheduling is attributed to the saving of freezing stack segments. Although our experiment confirms the good space performance of the lazy strategy, it gives a counterintuitive result that the eager strategy is as fast as the lazy strategy. This result implies that the cost of iterative evaluation is considerably smaller than that of freezing stack segments, and for predicates with cuts the eager strategy can be used without significant slow-down. In our tabling system, different answer consumption strategies can be used for different predicates. The tabling system described in \cite{Rocha05} also supports mixed strategies.

Semi-naive optimization is a fundamental idea for reducing redundancy in bottom-up evaluation of logic database queries \cite{Banc86,Ullman88}. As far as we know, its impact on top-down evaluation had been unknown before \cite{Zhou04}. OLDT \cite{Tamaki86} and SLG \cite{Sagonas98} do not need this technique since it is not iterative and the underlying delaying mechanism successfully avoids the repetition of any derivation step. An attempt has been made by Guo and Gupta \cite{Guo01} to make incremental consumption of tabled answers possible in DRA. In their scheme, answers are also divided into three regions but answers are consumed incrementally as in bottom-up evaluation. Since no condition is given for the completeness and no experimental result is reported on the impact of the technique, we are unable to give a detailed comparison.

Our semi-naive optimization differs from the bottom-up version in two major aspects: Firstly, no differentiated rules are used. In the bottom-up version differentiated rules are used to ensure that at least one new answer is involved in the join of answers for each rule. Consider, for example, the clause:
\begin{tabbing}
aa \= aaa \= aaa \= aaa \= aaa \= aaa \= aaa \kill
\> $H:-P,Q.$
\end{tabbing}
The following two differentiated rules are used in the evaluation instead of the original one:
\begin{tabbing}
aa \= aaa \= aaa \= aaa \= aaa \= aaa \= aaa \kill
\> $H:-\Delta P,Q.$ \\
\> $H:-P,\Delta Q.$
\end{tabbing}
Where $\Delta P$ denotes the new answers produced in the previous round for P. Using differentiated rules in top-down evaluation can cause considerable redundancy, especially when the body of a clause contains non-tabled subgoals.

The second major difference between our semi-naive optimization and the bottom-up version is that answers in our method are consumed sequentially until exhaustion, not incrementally as in bottom-up evaluation. A tabled subgoal consumes either all available answers or only new answers including answers produced in the current round. Neither incremental consumption nor sequential consumption seems satisfactory. Incremental consumption avoids redundant joins but may require more rounds to reach fixpoints. In contrast, sequential consumption never need more rounds to reach fixpoints but may cause redundant joins of answers. The early promotion technique alleviates the problem of sequential consumption. By promoting answers early from the {\it current} region to the {\it previous} region, we can considerably reduce the redundancy in joins.

Semi-naive optimization may lower time complexities in bottom-up evaluation \cite{Banc86}. The same result holds to the top-down version as demonstrated by Warren's example. Our experimental results show that semi-naive optimization gives an average speed-up of over $30\%$ to linear tabling if answers are promoted early, and almost no speed gain if no answer is promoted early. In linear tabling, only looping subgoals need to be iteratively evaluated. For non-looping subgoals, no re-evaluation is necessary and thus semi-naive optimization has no effect at all on the performance. Most of the looping subgoals in our chosen benchmarks reach their fixpoints after 2-3 iterations. In general, more iterations are needed to reach fixpoints in bottom-up evaluation. In addition, in bottom-up evaluation, the order of the joins can be optimized and no further joins are necessary once a participating set is known to be empty. In contrast, in linear tabling joins are done in strictly chronological order. For a conjunction $(P,Q,R)$, the join $P\Join Q$ is computed even if no answer is available for $R$. Because of all these factors, semi-naive optimization is not as effective in linear tabling as in bottom-up evaluation. 

Our semi-naive optimization requires the identification of last depending subgoals. For this purpose, a level mapping is used to represent the call graph of a given program. The use of a level mapping to identify optimizable subgoals is analogous to the idea used in the stratification-based methods for evaluating logic programs \cite{Apt88,Chen96,Przymusinski89}. In our level mapping, only predicate symbols are considered. It is expected that more accurate approximations can be achieved if arguments are considered as well.

Semi-naive optimization does not solve all the problems of recomputation in linear tabling. Recall the Warren's example:
\begin{verbatim}
   :-table p/2.
   p(X,Y) :- p(X,Z),c(Z,a,Y).
   p(X,Y) :- p(X,Z),c(Z,b,Y).
   p(X,X).
\end{verbatim}
Assume there is a very costly non-tabled subgoal preceding {\tt p(X,Z)}, then the subgoal has to be executed in each iteration even with semi-naive optimization. This example demonstrates the acuteness of the problem of recomputation because the number of iterations needed to reach the fixpoint is not constant. One treatment would be to table the subgoal to avoid recomputation, as suggested in \cite{Guo01}, but tabling extra predicates can cause other problems such as over consumption of table space.

\section{Conclusion}
In this paper we have described two answer consumption strategies (namely, {\it lazy} and {\it eager} strategies) and semi-naive optimization for linear tabling. We have compared the two strategies both qualitatively and quantitatively. Our results indicate that, while the lazy strategy has better space efficiency than the eager strategy, the eager strategy is comparable in speed with the lazy strategy. This result implies that for all-solution search programs the lazy strategy should be adopted and for partial-solution search programs including programs with cuts the eager strategy should be used.

We have tailored semi-naive optimization to linear tabling and have given sufficient conditions for it to be complete. Moreover, we have proposed a technique called {\it early answer promotion} to reduce redundant consumption of answers. Our experimental result indicates that semi-naive optimization gives significant speed-ups to some programs.

Linear tabling has several attractive advantages including its simplicity, ease of implementation, and good space efficiency. Early implementations of linear tabling were several times slower than XSB. This paper has demonstrated for the first time that linear tabling with optimization is as competitive as SLG on time efficiency as well for the benchmarks. 

Semi-naive optimization does not solve all the problems of recomputation in linear tabling. There are programs for which recomputation can be costly, even leading to higher complexities. The future work is to identify the patterns of such programs and find methods to deal with them.

\section{Acknowledgement}
The preliminary results of this article appear in ACM PPDP'03 and PPDP'04. Taisuke Sato is supported in part by CREST, and Yi-Dong Shen is supported in part by the National Natural Science Foundation of China grants numbered 60673103 and 60421001.

\end{document}